# Exceeding Conservative Limits: A Consolidated Analysis on Modern Hardware Margins

George Papadimitriou, *Member, IEEE*, Athanasios Chatzidimitriou, *Student Member, IEEE*,
Dimitris Gizopoulos, *Fellow, IEEE*, Vijay Janapa Reddi, *Member, IEEE*, Jingwen Leng,
Behzad Salami *Member, IEEE*, Osman S. Unsal *Member, IEEE*, and Adrian Cristal Kestelman, *Member, IEEE*

*Abstract*—**Modern large-scale computing systems (data centers, supercomputers, cloud and edge setups and high-end cyber-physical systems) employ heterogeneous architectures that consist of multicore CPUs, general-purpose many-core GPUs, and programmable FPGAs. The effective utilization of these architectures poses several challenges, among which a primary one is power consumption. Voltage reduction is one of the most efficient methods to reduce power consumption of a chip. With the galloping adoption of hardware accelerators (i.e., GPUs and FPGAs) in large datacenters and other large-scale computing infrastructures, a comprehensive evaluation of the safe voltage reduction levels for each different chip can be employed for efficient reduction of the total power. We present a survey of recent studies in voltage margins reduction at the system level for modern CPUs, GPUs and FPGAs. The pessimistic voltage guardbands inserted by the silicon vendors can be exploited in all devices for significant power savings. On average, voltage reduction can reach 12% in multicore CPUs, 20% in manycore GPUs and 39% in FPGAs.**

## I. INTRODUCTION

Process variations can affect the dimensions of the transistors (length, oxide thickness, etc.) due to the modern fabrication process, and thus, can impact the threshold voltage of a MOS device [1]. Such *static* variations remain constant after the release of the chip to the market. On top of that, transistor aging and *dynamic* variation in supply voltage and temperature, caused by different workload interactions can also affect the correct operation of a chip. Accounting the different types of variations, silicon vendors offset the best-case supply voltage with a fixed guardband to ensure reliability under worst-case conditions, as shown in Fig. 1a.

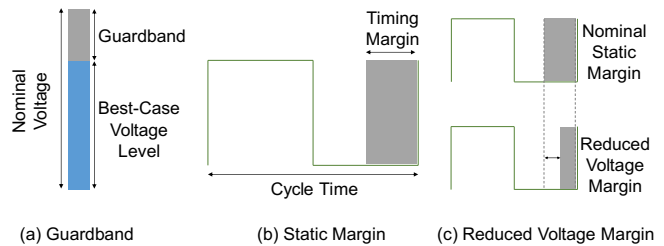

Fig. 1. Voltage guardband ensures reliability by effectively inserting extra timing margin. Reduced voltage margins can improve total system efficiency if they don't affect reliable operation.

The guardband results in faster circuit operation than required at the target frequency for typical workloads, which results in additional (thus wasted) cycle time, as shown in Fig. 1b. In case of a timing emergency caused by voltage droops, the extra margin prevents timing violations and failures by tolerating circuit slowdown. While static, worst-case guardbanding ensures robust execution for virtually all circumstances, it severely affects power and energy efficiency of the average case [2].

Supply voltage reduction (Fig. 1c) is one of the most efficient techniques to reduce the power consumption of the chip, because dynamic power is quadratic in voltage. Several system-level approaches have been proposed to predict and effectively utilize the safe operation limits (i.e., $V_{min}$) of the microprocessors. For example, the authors in [3] [4] propose an approach to predict the large voltage noise droops. Along the same lines the authors of [5] [6] proposed a firmware-based approach to predict the lowest safe voltage operation by observing corrected errors manifested on caches of an Intel Itanium processor. The energy gain in these studies comes from the variations of the $V_{min}$ when the same workload runs on different cores (core-to-core variation) or different workloads run on the same core (workload-to-workload variation).

Similar to multicore CPUs, many-core GPU architectures also require a large voltage guardband for reliable operation under all types of variations. However, their massive nature as well as their distinctive microarchitectural features render the traditional CPU-centric analysis framework and solutions unsuitable for GPUs. As such, prior work focuses on manycore-centric modeling, analysis, and smoothing techniques [23] [24] [25].

Furthermore, FPGAs are getting increasingly popular as acceleration platforms thanks to their massively parallel architecture and the capability of stream-fashion computation and data marshaling. Due to the development of High-Level Synthesis (HLS) tools in recent years, it is expected that FPGAs will be exploited in 1/3 of data centers by 2020 [15]. However, their power consumption is still a key concern, especially when compared against equivalent ASIC designs.

In this paper, we summarize recent system-level analysis and evaluation of safe voltage margins in multicore CPUs, manycore GPUs, and FPGAs. We aim to present consolidated results and observations for heterogenous architectures and summarize the emerging trends in hardware margins and energy-efficiency.

For the multicore CPUs part of this paper, we focus on two recent state-of-the-art ARMv8-compliant microprocessors. By experimenting on these recent multicore microprocessor chips, we present a number of observations which can potentially improve energy-efficiency in future designs. We report up to 18.4% power reduction in single-core executions and up to 17.6% in multicore CPU executions [8]-[14]; the multicore CPUs part of this survey was conducted in the context of the UniServer project [9] [14]. For the manycore GPUs part of this survey, we focus on a wide range of four NVIDIA GPUs spanning two generations (Fermi and Kepler). We report the $V_{min}$ points of different programs and we show as large as 20% voltage guardbands on these GPUs; the energy efficiency improvements through voltage reduction can be as large as 25% [24]-[28]. We also report the experimental evaluation of aggressive undervolting in FPGAs. By experimenting on real FPGA fabrics, we show the significant effectiveness of this technique to reduce their power consumption by on average 90%, first, by eliminating the voltage guardband which is measured by on average 39% and also by trading-off the power-reliability in further lower voltage levels [16]-[20]. The research on FPGA's undervolting is being conducted under LEGaTO project [21] [22]. Table I below presents the consolidated information about the voltage reduction and power savings potential for all the platforms of this study.

## II. EXCEEDING GUARDBANDS IN ARMv8 MULTICORE CPUs

### A. Applied Micro's X-Gene 2 and X-Gene 3 Specifications

Applied Micro's (now Ampere Computing) X-Gene 2 and X-Gene 3 multicore CPUs consist of 8 and 32 64-bit ARMv8-compliant cores, respectively. Both CPUs offer high-end processing performance. X-Gene 3 microprocessor has a main power domain that includes the CPU cores, the L1, L2 and L3 cache memories, and the memory controllers, which is called PCP (Processor ComPlex) power domain. X-Gene 2 has a similar structure; the difference is that it has 8 cores instead of 32, and the L3 cache, which is 8MB instead of 32MB, is located in a different power domain. The operating voltage of the main power domain can change from 980mV downwards in X-Gene 2, and from 870mV downwards in X-Gene 3. While all the CPU cores operate at the same voltage, each pair of cores (PMD – Processor MoDule) can operate at different frequency. Frequency ranges from 300MHz to 2.4GHz in X-Gene 2, and from 375MHz to 3GHz in X- Gene 3 (at 1/8 steps of the maximum clock frequency in both microprocessors).

### B. Exposing $V_{min}$ Values in Single-Core Executions

We experimentally obtain the $V_{min}$ values of 10 SPEC CPU2006 [7] benchmarks on the three X-Gene 2 chips (TTT, TFF, TSS) [8] [9] [10] [11] [12] [13], running the entire time-consuming undervolting experiments multiple times for each benchmark. This part of the study focuses on a quantitative analysis of the $V_{min}$ for diverse microprocessor chips of the same architecture in order to expose the potential guardbands of each chip, as well as to quantify how the program behavior affects the guardband and to measure the core-to-core and chip-to-chip variation. $V_{min}$ is defined as the minimal working voltage of the microprocessor for any workload or operating condition at a specific clock frequency.

For a significant number of benchmarks, we can see variations between different programs and different chips. Fig. 2 presents the most robust core for each chip, and for these programs the $V_{min}$ varies from 885mV to 865mV for TTT, from 885mV to 860mV for TFF, and from 900mV to 870mV for TSS. Considering that the nominal voltage for the X-Gene 2 is 980mV, there is a significant reduction of voltage without affecting the correct execution of programs (single-core runs), which is at least 9.7% for the TTT and TFF, and 8.2% for the TSS. The corresponding power (and corresponding energy) savings are 18.4% for the TTT and TFF chip, and 15.7% for the TSS chip. We also notice that the workload-to-workload variation (~3%) remains the same across the three chips of the same architecture; however, there is significant variation among the chips. This means that there is a program dependency of $V_{min}$ behavior in all chips.

TABLE I. MINIMUM/MAXIMUM VOLTAGE AND POWER REDUCTION FOR PUBLICLY AVAILABLE CPUs, GPUs, FPGAs STUDIES.

| Platform | | ISA / Family | Process Technology | $V_{dd}$ (mV) | Voltage Reduction | Power Reduction |
|---|---|---|---|---|---|---|
| X-Gene 2 | | ARMv8 | 28nm | 980 | 6.1% - 11.7% | 11.6% - 18.4% |
| X-Gene 3 | | ARMv8 | 16nm | 870 | 4.6% - 11.5% | 10.9% - 17.6% |
| Itanium | | IA-64 | 32nm | 1100 | 20% | 18% - 23% |
| Sandy Bridge-E | | x86-64 | 32nm | 1365 | 12.1% - 15% | 12% - 20% |
| Haswell | | x86-64 | 22nm | 844 | 9.95% - 10.55% | 17% - 20% |
| POWER7 | | POWER | 32nm | 1285 | 5.8%-7% | 13.8% average |
| GPU | Kepler | 28nm | 1090 | 9.2% - 18.3% | 8% - 25% |
| | Fermi | 40nm | 1090 | 11.6% - 20.3% | 7% - 22% |
| FPGA | VC707 | 28nm | 1000 | 41.2% | 89.2% |
| | ZC702 | 28nm | 1000 | 42.8% | 89.8% |
| | KC705 | 28nm | 1000 | 38.5% - 42.8% | 87.2% - 90.1% |

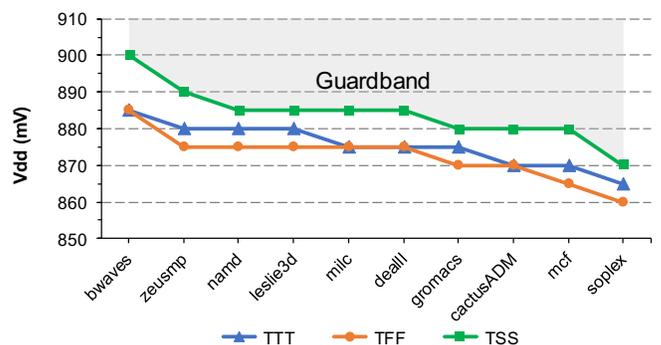

Fig. 2. $V_{min}$ single-core results at 2.4 GHz for 10 SPEC CPU2006 programs on 3 different X-Gene 2 chips (TTT, TFF, TSS).

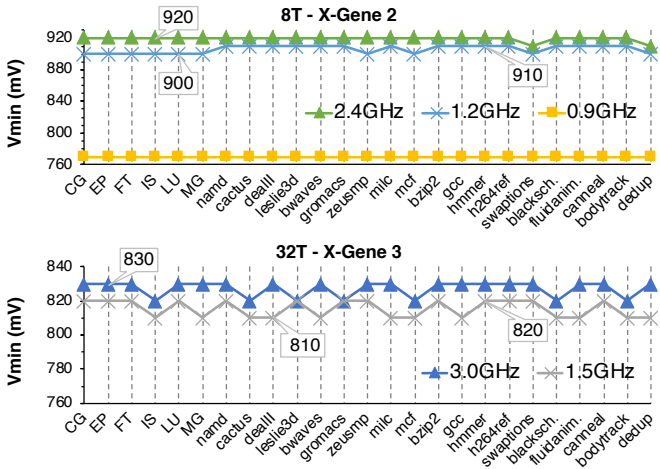

Fig. 3. $V_{min}$ results for multicore runs. The graph presents the X-Gene 2 $V_{min}$ points (top) with 8 threads in 2.4 GHz, 1.2 GHz, and 0.9 GHz clock frequencies, and the X-Gene 3 $V_{min}$ points (bottom) with 32 threads in 3 GHz and 1.5 GHz clock frequencies.

### C. Exposing $V_{min}$ Values in Multi-Core Executions

Through massive characterization experiments running 25 multi-threaded benchmarks, we obtained the multicore $V_{min}$ values on the two different technology ARMv8-compliant microprocessors: X-Gene 2 and X-Gene 3 (28nm and 16nm, respectively). Fig. 3 shows the $V_{min}$ characterization results for the 25 benchmarks on X-Gene 2 with 8-thread executions of the benchmarks for the three different frequencies: 2.4GHz, 1.2GHz, 0.9GHz, and X-Gene 3 with 32-thread executions for 3GHz and 1.5GHz, respectively [14]. Fig. 3 shows, that for the same number of threads and at the same frequency, the $V_{min}$ for all 25 benchmarks is virtually the same. There are some cases, where a benchmark has a little lower $V_{min}$, only 10mV or ~1% of the nominal voltage.

To understand this phenomenon, we study the voltage droop magnitude of the microprocessors for all the different frequency and core allocation configurations, by leveraging the embedded oscilloscope in the X-Gene 3 microprocessor. Fig. 4 presents two different ranges of voltage droop magnitude when the microprocessor operates at 3GHz: (a) the [55mV, 65mV) in which we present the configurations of all programs that produce voltage droops more than or equal to 55mV and less than 65mV, and (b) the [45mV, 55mV) in which we present the configurations of all programs that produce voltage droops more than or equal to 45mV and less than 55mV.

As we can see on the left graph of Fig. 4, the configurations with 32 threads and 16 spreaded threads (one thread is running on each PMD), which means that all 16 PMDs of the microprocessor running at 3GHz frequency produce voltage droop magnitude between 55mV and 65mV. However, the configuration of 16 clustered threads (two threads running on a PMD) has almost zero droops in the range of [55mV, 65mV) for all programs. On the right graph of Fig. 4, the configurations with 16 clustered threads and 8 spreaded threads produce voltage droop magnitude in the range of [45mV, 55mV). Thus, the emergency voltage droops are massive and lead to virtually workload-independent $V_{min}$.

Although the workload variability marginally affects the $V_{min}$ in multicore executions, core allocation and clock frequency are the major contributors to the $V_{min}$. The reason is that the frequency and different core allocations are the main factors that can affect the emergency voltage droop magnitude. In particular, the largest amount of voltage reduction (12%) is a result of clock division in a specific clock frequency, while just one step further frequency reduction (due to clock skipping) delivers 3% further voltage reduction. Moreover, assigning the running threads in different cores, we can achieve up to 3% more voltage reduction.

Combining all observations for single-core and multicore characterization, we obtain an optimal scheme of the microprocessors when running real workloads, which can achieve on average 25.2% energy savings on X-Gene 2, and 22.3% energy savings on X-Gene 3, with a minimal performance penalty of 3.2% on X-Gene 2 and 2.5% on X-Gene 3 compared to the default voltage and frequency microprocessor's conditions.

### III. VOLTAGE REDUCTION RESULTS FOR x86-64 MICROPROCESSORS

In this part of the study, we compare two x86-64 microprocessors (a Sandy Bridge-E and a Haswell) concerning their safe voltage guardbands by running the benchmarks in each individual core of the two chips [53].

In Fig. 5, we illustrate the percentage of voltage reduction from the nominal voltage for all benchmarks of our study that run on each of the 6 cores of the Sandy Bridge-E, and of each of the 2 cores of the Haswell microprocessor. In general, the Sandy Bridge-E has higher percentage of voltage reduction within a range between 12.09% and 15.02%, while the same percentage for the Haswell ranges from 9.95% to 10.55%. The highest variation is observed in Core 5 of the Sandy Bridge-E microprocessor, which shows a voltage reduction between 12.09% and 15.02%.

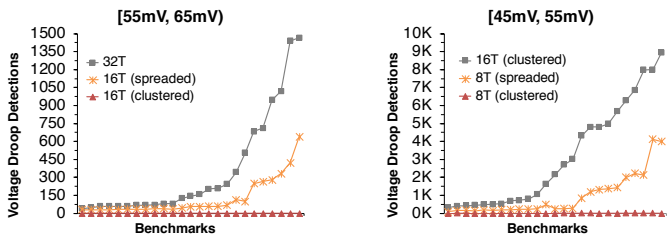

Fig. 4. Voltage droop detections for each program in 2 different voltage droop magnitudes.

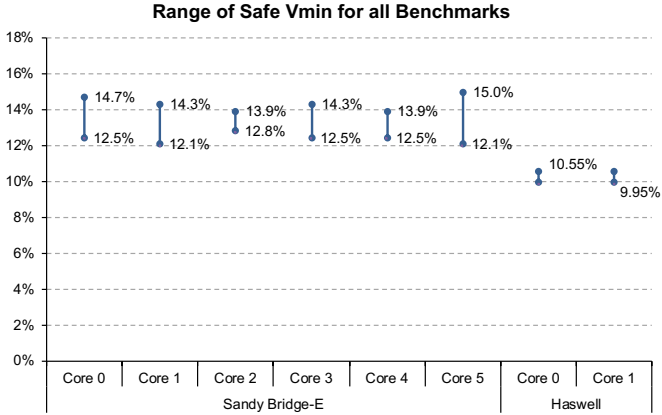

Fig. 5. Range of safe voltage reduction for Sandy Bridge-E and Haswell microprocessors [53].

## IV. MANY CORE GPUs

### A. Opportunity for Guardband Optimization

The margin between the nominal voltage and $V_{min}$ reflects the optimization potential. Measurement results in Fig. 6a show that GPUs require as large as 20% voltage guardbands to tolerate worst-case conditions. Fig. 6b shows that measured energy efficiency improvements of reducing the voltage guardband can be as large as 25% [26] [28].

The minimum energy saving is only 8%, which suggests the guardband optimization potential is program-dependent. Voltage guardband is often impacted by different variation types, including process, voltage, and thermal (PVT) variations. The work [26] uses the method of exclusion to rule out other types of variations and identify the voltage noise as the major consumer of voltage guardband and also the root cause of program dependent guardband behavior. The work reports both process and temperature variation have a relatively uniform impact on the $V_{min}$ across all programs. The process variation consumes about 0.07V margin while the temperature variation consumes about 0.02V margin. This leaves 0.1V margin for voltage noise.

### B. Modeling

To understand voltage guardbands in more detail, we need a modeling framework. *GPUVolt* [30] simulates the voltage noise behavior by calculating the time domain response of the power (voltage) delivery model under current input profiles of each core (Fig. 7). We use *GPUWattch* [27], a cycle-level GPU power simulator, to approximate the current variation profile of each GPU core under a certain supply voltage level. *GPUWattch* takes the microarchitectural activity statistics from *GPGPU-Sim* [31], a cycle-level simulator, and calculates the power consumption of each microarchitectural component.

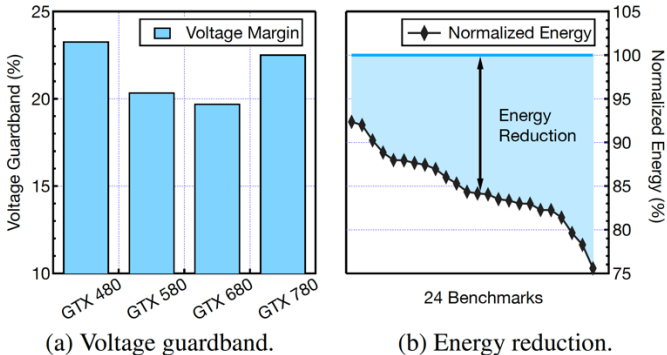

| (a) Voltage guardband. | (b) Energy reduction. |

Fig. 6. Opportunity of exploiting the voltage guardband. (a) Measured voltage guardband on four commercial graphics cards. (b) Measured energy reduction on a GTX 480.

*GPUVolt*'s power delivery model consists of three parts (Fig. 7): the printed circuit board (PCB), the package, and the on-die power delivery network (PDN). We use a lumped model for the PCB and package circuit and a distributed model for the on-die PDN. The distributed model can reflect both intra-SM as well as inter-SM voltage noise interference. We also propose a TPD based PDN scaling methodology since there is no public information on its actual PDN design. The validation results show *GPUVolt* has a Pearson's correlation of 0.9 with hardware based $V_{min}$ measurement.

### C. Root Cause

We characterize and analyze the root cause for the large voltage droops in the manycore architecture using *GPUVolt*. We perform the analysis at the single-core level to study the impact of individual microarchitectural component and then enlarge the scope of the analysis to the entire processor level to study the inter-core voltage interference.

**Single-Core.** We leverage the linear property of the voltage model to quantify each component's contribution to a single SM's voltage noise. Fig. 8 shows the contribution of the major components. The register file is the single most dominant source of voltage droops, which is closely tied to the GPU's unique characteristics. Modern GPUs require a large register file to hold the architectural states of thousands of concurrent threads (multiple times of the L1 cache size). Consequently, the GPU's RF access rate and power consumption are much higher compared to the CPU [27] [32].

**Many Core.** To understand the voltage noise characteristics in manycore GPUs, we propose a conceptual spatial-temporal characterization framework [29]. In short, it examines voltage noise in the temporal (i.e., time varying) and spatial (i.e., core versus chip-wide) dimensions. Using this framework, we determined that there are two main types of GPU voltage noise: the fast-occurring first-order droops that are localized to a small cluster of neighboring cores, and the slow-occurring chip-wide second-order droops, shown in Fig. 9.

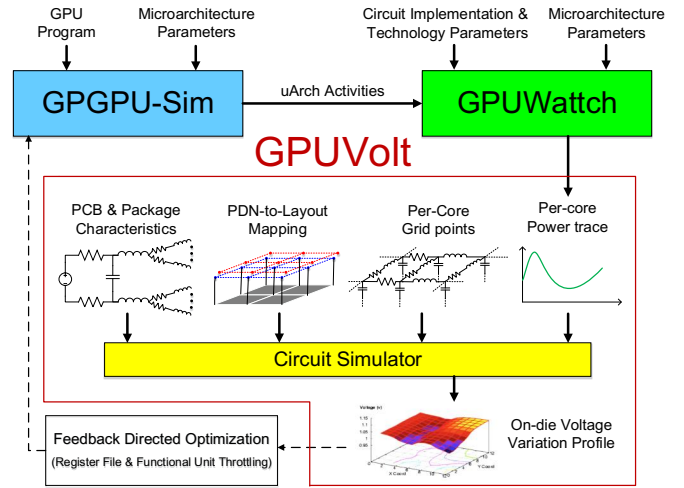

Fig. 7. Overview of the integrated performance, power, and voltage modelling framework.

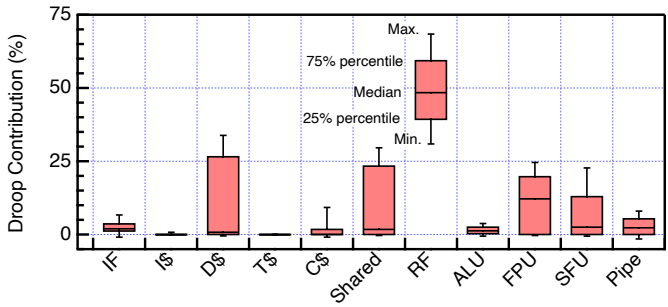

Fig. 8. Component contribution to any voltage droop greater than 3% at the single-SM level.

We identify the sensitivity to the activity alignment as the reason why a particular droop type is present/absent. The first-order droop occurs very fast and requires almost perfect alignment for multiple cores to resonate (i.e., global droop). In contrast, the second-order droop occurs much slower and there exists loosely aligned core activity owing to GPU's single-program multiple-data execution model (which we call implicit synchronization) that can cause global droops. These events that can lead to implicit synchronization include I-/D-cache miss and thread block launch.

### D. Voltage Guardband Optimization

We have studied two types of voltage guardband optimization. The first optimization, voltage smoothing, mitigates the worst-case voltage droop magnitude and therefore enables a tighter operating margin. The second optimization, predictive guardbanding, dynamically adapts to the voltage fluctuation for larger energy saving.

**Smoothing.** To smooth GPU voltage noise, we introduce *hierarchical voltage smoothing* [29], where each level specifically targets one type of voltage droop. For the first-order droop, we train a prediction model (off-line trained) to predict the local first-order droop using the root-cause analysis data based on register file and dispatch unit activity. The models provide enough response time for smoothing to work. For the second-order droop, caused by the implicit synchronization, the smoothing mechanism leverages existing hardware communication mechanisms and delays execution to disrupt the current and future synchronization pattern. The hierarchical mechanism reduces the worst-case droop by 31%, which enables a smaller voltage guardband for energy-efficiency improvements. We observe an average 7.8% savings.

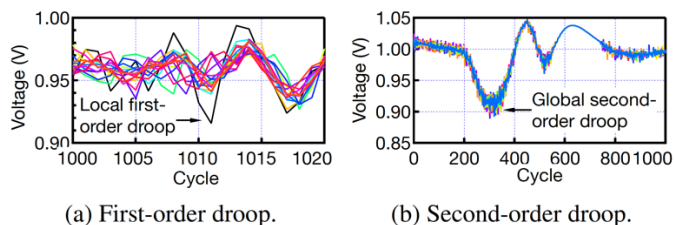

(a) First-order droop.     (b) Second-order droop.

Fig. 9. Two major voltage droop types in GPUs.

**Adaptation.** Besides the minimum voltage reduction, we also proposed a novel dynamic voltage adaptation approach for more energy savings. We show that we can use a kernel' microarchitectural performance counters to predict its $V_{min}$ value accurately [26]. Based on the $V_{min}$ prediction, we propose to use the CPU to manage the GPU's guardband at the granularity of kernel, which is the minimal control and scheduling unit by the CPU. Instead of using the complicated feedback loop for margin adaption as in the Power 7 CPU [34], our scheme collects each kernel's performance counters and uses the $V_{min}$ predictor to reduce the operating margin. Moreover, it is unlikely that managing the guardband at a finer granularity could bring more benefits. The reason is the highly possible recurring pattern of large voltage droops during a kernel execution. Although we cannot measure it from the hardware, prior simulation-based study has shown that large voltage droops frequently occur during the kernel execution because of the throughput-optimized GPU architecture. If large voltage droops happen frequently, managing the voltage with a finer granularity is unlikely to lead to more benefits.

Although our model predicts $V_{min}$ within small (3%) error margin for training programs, the prediction error for unseen programs, with different characteristics from the training programs, could exceed this error margin. This kind of corner case can result in a system failure, which makes the safety net necessary in the predictive guardbanding. In our work, we recognize the fact that the GPU is a coprocessor and propose to offload the GPU error recovery to the CPU, which forms our asymmetric resilience fail-safe mechanism [33] [57]. Overall, the predictive guardbanding realizes the 20% energy saving.

## V. SUPPLY VOLTAGE REDUCTION IN FPGAS

### A. Experimental Results

Experiments are performed on representative FPGAs from Xilinx, a main vendor, including VC707 (performance-oriented Virtex), two identical samples of KC705 (A & B, power-oriented Kintex), and ZC702 (CPU-based Zynq). Among various FPGA components, a major part of experiments is initially performed on on-chip memories or Block RAMs (BRAMs), thanks to their importance in the architecture of state-of-the-art applications like FPGA-based DNNs as well as the capability of their voltage rail to be independently regulated. BRAMs are a set of small blocks of SRAMs, distributed over the chip, and in a programmable-fashion can be chained to build larger memories. All evaluated platforms are fabricated with 28nm technology and their nominal/default BRAM's voltage level (VCCBRAM) is 1V.

As shown in Fig. 10, undervolting VCCBRAM below the nominal level, the performance or reliability of the BRAMs are not affected until a certain level, *i.e.*, minimum safe voltage or $V_{min}$. This region is the GUARDBAND, which is mainly considered by vendors to ensure the worst-case environmental and process scenarios. In the GUARDBAND voltage region, data can be safely retrieved without compromising reliability. Further undervolting, although the

FPGA is still accessible, the content of some BRAMs experience faults or bit-flips. We call it as the CRITICAL region. Finally, further undervolting, the DONE pin is unset at Vcrash and the FPGA does not respond for any request in the CRASH region. As seen, there is a slight difference of mentioned voltage margins among platforms even for identical samples of KC705; however, those three voltage regions are recognizable for all.

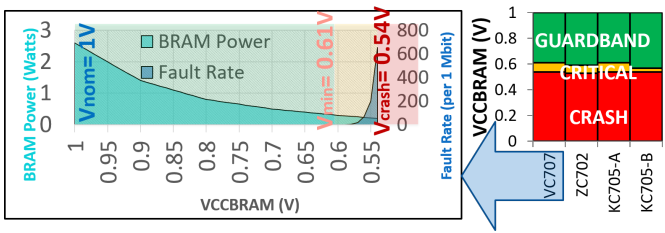

Fig. 10. Voltage behavior and power/reliability trade-off behavior of FPGAs (@ambient temperature).

As shown in Fig. 10 (for VC707), the power is continuously reduced through undervolting in both GUARDBAND and CRITICAL voltage regions; however, within the CRITICAL region, some of the memories are infected. The fault rate exponentially increases by further undervolting within the CRITICAL region and arrives to 652 faults/Mbit at Vcrash. In the same line, we observe that the fault rate exponentially increases up to 153, 254, and 60 faults/Mbit at Vcrash for ZC702, KC705-A, and KC705-B, respectively.

### B. Case Study: Deep Neural Network (DNN)

DNNs applications are widely used in the context of nonlinear cognitive problems. However, they have a power- and compute-intensive nature. Hence, the aggressive undervolting technique briefly described above, can potentially improve the energy-efficiency of such DNN models implemented on FPGAs. Toward this goal, we develop a typical model of FPGA-based DNN in the inference phase, as detailed in [16]. In the evaluated model, weights of DNN are located in FPGA BRAMs; hence, $V_{CCBRAM}$ undervolting can cause bit flips in DNN weights which can subsequently result in accuracy loss in the classification task of the DNN. Fig. 11 depicts the experimental results of this study in terms of the accuracy loss and power-efficiency achievement trade-off. As seen, we achieve a significant power saving of more than 90% by eliminating the voltage guardband and an additional saving of 60% by further undervolting below the $V_{min}$. However, due to the undervolting faults at supply voltage levels below $V_{min}$, the DNN accuracy experiences a considerable drop of more than 3%. To alleviate this issue, we present and evaluate effective

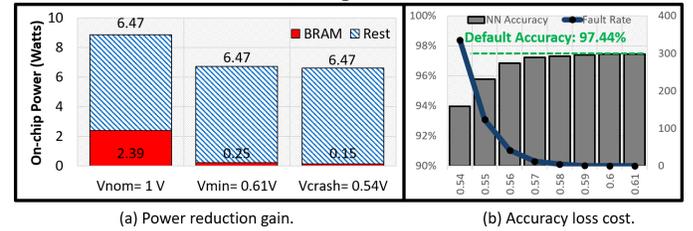

Fig. 11. Power and accuracy loss trade-off in FPGA-based DNN through aggressive undervolting on $V_{CCBRAM}$. Experiments are for the DNN benchmark of MNIST on VC707.

fault mitigation techniques that are described in detail in [16] and [18].

### C. Environmental Temperature

We extend the experimental study to evaluate the impact of the environmental temperature on the reliability of FPGA. Toward this goal, we place the FPGA boards inside a heat chamber where the environmental temperature can be regulated. As can be seen in Fig. 12, with heating up, the fault rate constantly reduces; for instance, by more than 3X for 30°C higher temperature (for VC707). Also, the changes on the fault rate over the voltage are significantly different among platforms evaluated; for instance, as can be seen, a relatively 156% more fault rate at 50°C is reduced to 11.6% less fault rate at 80°C, for VC707 vs. KC705-A. The significant variation on the fault rate of different platforms and the impact of the temperature can be the consequence of the architectural and technological difference, process variation, or aging effects among them.

## VI. RELATED WORK

During the last years, the goal for improving microprocessors' energy efficiency, while reducing their power supply voltage is a major concern of many scientific studies that investigate the chips' operation limits in nominal and off-nominal conditions. In this section, we briefly summarize the existing studies and findings concerning low-voltage operation and characterization studies.

Wilkerson et al. [47] go through the physical effects of low-voltage supply on SRAM cells and the types of failures that may occur. After describing how each cell has a minimum operating voltage, they demonstrate how typical error protection solutions start failing far earlier than a low-voltage target (set to 500mV) and propose two architectural schemes for cache memories that allow operation below 500 mV. The word-disable and bit-fix schemes sacrifice cache capacity to tolerate the high failure rates of low voltage operation. While both schemes use the entire cache on high voltage, they sacrifice 50% and 25% accordingly in 500 mV. Compared to existing techniques, the two schemes allow a 40% voltage reduction with power savings of 85%.

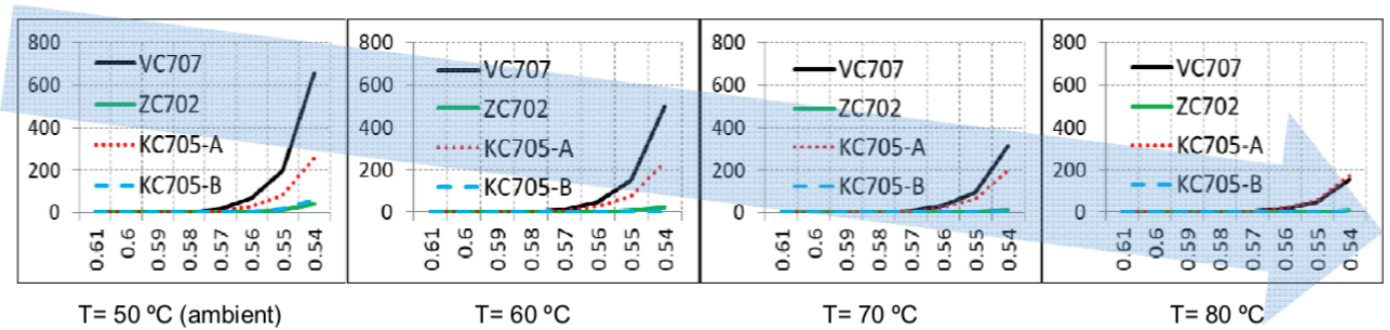

Fig. 12. Reliability of low-voltage BRAMs by experimenting with the environmental temperature changes (x-axis: $V_{CCBRAM}$, y-axis: fault rate).

Chishti et al. [48] propose an adaptive technique to increase reliability of cache memories, allowing high tolerance on multi-bit failures that appear on low-voltage operation. The technique sacrifices memory capacity to increase the error-correction capabilities, but unlike previously proposed techniques, it also offers soft and non-persistent error tolerance. Additionally, it does not require self-testing to identify erratic cells in order to isolate them. The MS-ECC design can achieve a 30% supply voltage reduction with 71% power savings and allows configurable ECC capacity by the operating system based on the desired reliability level.

Bacha et al. [5] present a new mechanism for dynamic reduction of voltage margins without reducing the operating frequency. The proposed mechanism does not require additional hardware as it uses existing error correction mechanisms on the chip. By reading their error correction reports, it manages to reduce the operating voltage while keeping the system in safe operation conditions. It covers both core-to-core and dynamic variability caused by the running workload. The proposed solution was prototyped on an Intel Itanium 9560 processor and was tested using SPECjbb2005 and SPEC CPU2000-based workloads. The results report promising power savings that range between 18% to 23%, with marginal performance overheads.

Bacha et al. [6] again rely on error correction mechanisms to reduce operating voltage. Based on the observation that low-voltage errors are deterministic, the paper proposes a hardware mechanism that continuously probes weak cache lines to fine-tune the system's supply voltage. Following an initial calibration test that reveals the weak lines, the mechanism generates simple write-read requests to trigger error-correction and is capable to adapt to voltage noise as well. The proposed mechanism was implemented as proof-of-concept using dedicated firmware that resembles the hardware operation on an Itanium-based server. The solution reports an average of 18% supply voltage reduction and an average of 33% power consumption savings, using a mix set of applications.

Bacha et al. [50] exploit the observation of deterministic error distribution to provide physically unclonable functions (PUF) to support security applications. They use the error distribution of the lowest save voltage supply as an unclonable fingerprint, without the typical requirement of additional dedicated hardware for this purpose. The proposed PUF design offers a low-cost solution for existing processors. The design is reported to be highly tolerant to environmental noise (up to 142%) while maintaining very small misidentification rates (bellow 1ppm). The design was tested on a real system using Itanium processor as well as on simulations. While this study serves a different domain, it highlights the deterministic error behavior on SRAM cells.

Duwe et al. [49] propose an error-pattern transformation scheme that re-arranges erratic bit cells that correspond to uncorrectable error patterns (e.g., beyond the correctable capacity) to correctable error patterns. The proposed method is low-latency and allows the supply voltage to be scaled further that it was previously possible. The adaptive rearranging is guided using the fault patterns detected by self-test. The proposed methodology can reduce the power consumption up to 25.7%, based on simulated modeling that relies on literature SRAM failure probabilities.

There are several papers that explore methods to eliminate the effects of voltage noise. Gupta et al. [39] and Reddi et al. [3] focus on the prediction of critical parts of benchmarks, in which large voltage noise glitches are likely to occur, leading to malfunctions. In the same context, several studies were presented to mitigate the effects of voltage noise [38] [39] [40] [41] [42] or to recover from them after their occurrence [43]. For example, in [35] [36] [37] the authors propose methods to maximize voltage droops in single core and multicore chips in order to investigate their worst-case behavior due to the generated voltage noise effects.

Similarly, authors in [51] and [52] proposed a novel methodology for generating dI/dt viruses that is based on maximizing the CPU emitted electromagnetic (EM) emanations. Particularly, they have shown that a genetic algorithm (GA) optimization search for instruction sequences that maximize EM emanations and generates a dI/dt virus that maximizes voltage-noise. They have also successfully applied this approach on 3 different CPUs: two ARM-based mobile CPUs and one AMD Desktop CPU [45] [46].

Lefurgy et al. [34] propose the adaptive guardbanding in IBM Power 7 CPU. It relies on the critical path monitor (CPM) to detect timing margin. It uses a fast CPM-DPLL (digital phase lock loop) control loop to avoid possible timing failures: when the detected margin is low, the fast loop quickly stretches the clock. To mitigate the possible frequency loss, adaptive guardbanding also uses a slow loop to boost the voltage when the averaged clock frequency is below the target. Leng et al. [26] studies the voltage guardband on the real GPU and shows the majority of GPU voltage margin protects against voltage noise. To fulfil the energy saving in the guardband, the authors propose to manage the GPU voltage margin at the kernel granularity. They study the feasibility of using a kernel's performance counters to predict the $V_{min}$, which enables a simpler predictive guardbanding design for GPU-like co-processors.

Aggressive voltage underscaling has been recently in part for FPGAs, as well. Ahmed et al. [44] extend a previously proposed offline calibration-based DVS approach to enable DVS for FPGAs with BRAMs using a testing circuitry to ensure that all used BRAM cells operate safely while scaling the supply voltage. L. Shen et al. [54] propose a DVS technique for FPGAs with Fmax; however, voltage underscaling below the safe level is not thoroughly investigated. Ahmed *et al.* [55] evaluate and compare the voltage behavior of different FPGA components such as LUTs and routing resources and design FPGA circuitry that is better suited for voltage scaling. Salamat *et al.* [56] evaluate a couple of FPGA-based DNN accelerators with low-voltage operations; however, using simulators.

## VII. Conclusion

In this paper, we summarize system-level evaluations of the voltage margins of recent multicore CPUs, manycore GPUs, and FPGAs. We first presented the results on two recent state-of-the-art ARMv8-compliant microprocessors chips, and two different Intel microprocessors. By leveraging the pessimistic voltage guardbands on the ARMv8 microprocessors, we can achieve up to 18.4% power reduction in single-core executions and up to 17.6% in multicore executions. Similarly, on Intel microprocessors we can achieve up to 20% power reduction for different workloads. For the manycore GPUs case, we report a comprehensive measurement using four NVIDIA GPUs (Fermi and Kepler architectures). We showed that manycore GPUs require as large as 20% voltage guardbands, while the energy efficiency improvements of reducing this guardband can be as large as 25%. We also reported experimental evaluation of aggressive undervolting in FPGAs, and showed a power consumption reduction of 90% on average more specifically for FPGA on-chip memories, (on average 39% voltage reduction).

## VIII. Acknowledgment

The research leading to results in the CPU part was funded by the EU H2020 Programme under the UniServer project (http://www.uniserver2020.eu), grant agreement nₒ 688540. Also, the research leading to results in FPGA part was funded by the EU H2020 Programme under the LEGaTO project (https://legato-project.eu), grant agreement nₒ 780681.

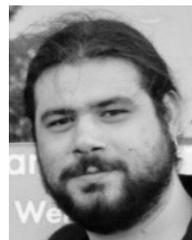
**George Papadimitriou** (S'14, M'19) is a Postdoctoral Researcher at the Dept. of Informatics & Telecomm. of the University of Athens. He graduated from the Dept. of Computer Systems Engineering of the Technological Educational Institute of Piraeus as valedictorian in 2011. He received his MSc with honors in Computer Systems Technology and his PhD in Computer Architecture from the Dept. of Informatics & Telecom. of the University of Athens. His research focuses on dependability and energy-efficient computer architectures, microprocessor reliability, functional correctness of hardware designs and design validation of microprocessors and microprocessor-based systems, in which he has published more than 20 papers in international conferences and journals.

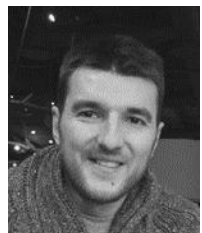
**Athanasios Chatzidimitriou** received his B.Sc. degree on Informatics engineering from Technological Educational Institute of Athens, Greece, his M.Sc. degree on Embedded Computing Systems from University of Piraeus, Greece and his Ph.D. on Computer Architecture from Department of Informatics and Telecommunications at University of Athens. His research interests focus on microprocessor reliability where he has published more than 25 papers in international conferences and journals.

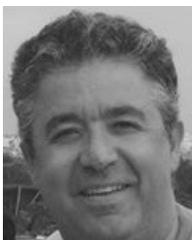
**Dimitris Gizopoulos** (S'93-M'97-SM'03 -F'13) is professor at the Department of Informatics & Telecommunications, University of Athens where he leads the Computer Architecture Lab. His research focuses on the dependability, performance and energy of computer architectures and systems. Gizopoulos has published more than 180 papers in peer reviewed IEEE and ACM journals and conferences and is author and editor of several books on dependable computing. He has served as associate editor and special issues guest editor for several Transactions and

Magazines (currently on the Editorial Board of IEEE Transactions on Computers, IEEE Transactions on Sustainable Computing, IEEE Transactions on Emerging Topics in Computing and ACM Computing Surveys) as well as the Organizing, Program and Steering committees of international events (currently General Chair of MICRO 2020 and Program Chair of IOLTS 2020). He is an IEEE Fellow, a Golden Core Member of IEEE Computer Society and an ACM Senior Member.

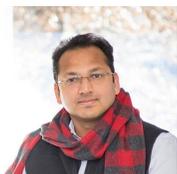

**Vijay Janapa Reddi** is an Associate Professor in John A. Paulson School of Engineering and Applied Sciences at Harvard University. Prior to that, he was an Associate Professor at The University of Texas at Austin in the Department of Electrical and Computer Engineering. His research interests include computer architecture and runtime systems, specifically in the context of autonomous machines and mobile and edge computing systems. He received a Ph.D. in computer science from Harvard University.

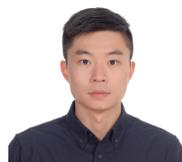

**Jingwen Leng** is a tenure-track Associate Professor in the John Hopcroft Computer Science Center and Computer Science Department at Shanghai Jiao Tong University. He received his Ph.D. from the University of Texas at Austin, where he focused on improving the efficiency and resiliency of general-purpose GPUs. He is currently interested at taking a holistic approach to optimizing the performance, efficiency, and reliability for heterogeneous computing systems.

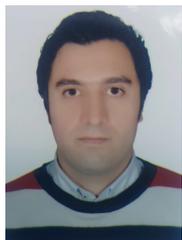

**Behzad Salami** is a post-doctoral researcher in the Computer Science (CS) department of Barcelona Supercomputing Center (BSC). He received his Ph.D. with honors in Computer Architecture from Universitat Politècnica de Catalunya (UPC) in 2018. Also, he obtained MSc and BSc degrees in Computer Engineering from Amirkabir University of Technology (AUT) and Iran University of Science and Technology (IUST), respectively. His research interests are heterogeneous computing and low-power & fault-resilient hardware accelerators.

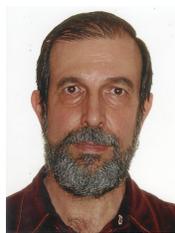

**Osman Sabri Unsal** is co-manager of the Parallel Paradigms for Computer Architecture research group at Barcelona Supercomputing Center (BSC). He got his B.S., M.S. and PhD in Computer Engineering from Istanbul Technical University, Brown University and University of Massachusetts, Amherst respectively. His current research interests are in computer architecture, fault-tolerance, energy-efficiency and heterogeneous computing. He is currently leading the LEGaTO EU H2020 research project on heterogeneous energy-efficient computing.

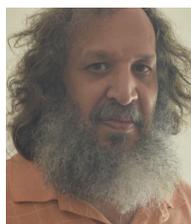

**Adrian Cristal** received the Licenciatura degree in computer science from the Faculty of Exact and Natural Sciences, Universidad de Buenos Aires, Buenos Aires, Argentina, in 1995, and the Ph.D. degree in computer science from the Universitat Politécnica de Catalunya (UPC), Barcelona, Spain. Since 2006, he is a Co-Manager of the Computer Architecture for Parallel Paradigms Research Group at Barcelona supercomputing Center. His current research interests include the areas of microarchitecture, multicore and heterogeneous architectures, and programming models for multicore architectures. Currently he is leading the architecture development of the vector processor unit in the European Processor Initiative.